\documentclass[conference]{IEEEtran}
\IEEEoverridecommandlockouts
\usepackage{cite}
\usepackage{amsmath,amssymb,amsfonts}
\usepackage{graphicx}
\usepackage{textcomp}
\usepackage{xcolor}
\usepackage{multirow}
\usepackage{cleveref}
\usepackage{float}

\def\BibTeX{{\rm B\kern-.05em{\sc i\kern-.025em b}\kern-.08em
    T\kern-.1667em\lower.7ex\hbox{E}\kern-.125emX}}

\usepackage{algpseudocode}
\usepackage[ruled,vlined]{algorithm}
\usepackage{graphicx}
\usepackage{subcaption}
\usepackage{comment}
\usepackage{tabularx} 
\usepackage{fancyvrb}
\usepackage{mdframed}
\usepackage{xcolor}
\usepackage[letterpaper, top=0.76in, bottom=1.1in, left=0.63in, right=0.63in]{geometry}

\begin{document}

\title{Lightweight LLMs for Network Attack Detection in IoT Networks\\
\thanks{This paper was accepted and presented at the 7th Computing, Communications and IoT Applications Conference (ComComAp 2025), held in Madrid, Spain, during 14-17 December 2025.}
}

\author{
    \IEEEauthorblockN{
        Piyumi Bhagya Sudasinghe\IEEEauthorrefmark{1},
        Kushan Sudheera Kalupahana Liyanage\IEEEauthorrefmark{1},
        Harsha S. Gardiyawasam Pussewalage\IEEEauthorrefmark{2}
    }
    \IEEEauthorblockA{
        \IEEEauthorrefmark{1}Department of Electrical and Information Engineering, University of Ruhuna, 81000 Matara, Sri Lanka\\
        \IEEEauthorrefmark{2}Department of Information and Communication Technology, University of Agder (UiA), N-4898 Grimstad, Norway\\
        Email: \{piyumi.s, kushan\}@eie.ruh.ac.lk; harsha.sandaruwan@uia.no
    }
}


\maketitle
\thispagestyle{empty}

\begin{abstract}
The rapid growth of Internet of Things (IoT) devices has increased the scale and diversity of cyberattacks, exposing limitations in traditional intrusion detection systems. Classical machine learning (ML) models such as Random Forest and Support Vector Machine perform well on known attacks but require retraining to detect unseen or zero-day threats. This study investigates lightweight decoder-only Large Language Models (LLMs) for IoT attack detection by integrating structured-to-text conversion, Quantized Low-Rank Adaptation (QLoRA) fine-tuning, and Retrieval-Augmented Generation (RAG). Network traffic features are transformed into compact natural-language prompts, enabling efficient adaptation under constrained hardware. Experiments on the CICIoT2023 dataset show that a QLoRA-tuned LLaMA-1B model achieves an F1-score of 0.7124, comparable to the Random Forest (RF) baseline (0.7159) for known attacks. With RAG, the system attains 42.63\% accuracy on unseen attack types without additional training, demonstrating practical zero-shot capability. These results highlight the potential of retrieval-enhanced lightweight LLMs as adaptable and resource-efficient solutions for next-generation IoT intrusion detection.

\end{abstract}


\section{Introduction}

The rapid proliferation of IoT and Industrial Control System (ICS) devices has substantially expanded the attack surface of modern networks~\cite{zong2025integrating}. As cyber threats grow in sophistication, robust network attack detection remains a critical component of cybersecurity infrastructure~\cite{panchal2024survey}. Traditional ML approaches, such as Random Forest (RF) and Support Vector Machine (SVM), achieve high accuracy on small-scale or binary intrusion detection tasks~\cite{chang2017network,hiremath2025ml}, but their performance deteriorates in realistic multi-class IoT environments with heterogeneous attack types~\cite{disha2022giwrf}. Moreover, these supervised models cannot effectively detect previously unseen (zero-day) attacks without retraining, limiting their adaptability to evolving threats.

Recent advances in Deep Learning (DL) and natural language processing (NLP) have introduced Large Language Models (LLMs) as versatile tools capable of modeling contextual dependencies across diverse data modalities~\cite{kaur2025harnessing}. Transformer-based LLMs have demonstrated strong potential for intrusion detection tasks by leveraging semantic understanding and sequence reasoning~\cite{kaur2025harnessing, zhang2025malicious}. Parameter-efficient fine-tuning methods such as Low-Rank Adaptation (LoRA) and Quantizes (QLoRA) further enable domain-specific adaptation with minimal computational and memory overhead, making them well-suited for IoT-scale deployments~\cite{mao2025lora}.

Transformer architectures are generally categorized as encoder-only, decoder-only, or encoder–decoder models. Encoder-based models (e.g., BERT, RoBERTa) are effective for classification and structured data representation, while decoder-only models (e.g., GPT-2, LLaMA, Mistral) are optimized for autoregressive text generation. Decoder models offer distinct advantages in reasoning, interpretability, and zero-shot adaptation, particularly when coupled with Retrieval-Augmented Generation (RAG)~\cite{alhammouri2025_hybrid_iot,loumachi2025gendfir}.

In this work, we present a unified LLM-based framework that handles both known attacks via QLoRA fine-tuning and unknown attacks via RAG, overcoming the retraining limitations of traditional ML. Using the CICIoT2023 dataset~\cite{hammal2023ciciot}, we conduct two experiments: (1) supervised fine-tuning of lightweight decoder-only LLMs for known attack detection, and (2) zero-shot detection of unseen attacks leveraging RAG.
The main contributions of this work are:
\begin{itemize}
\item \textbf{LLM-based detection:} Adapting lightweight decoder LLMs with QLoRA for multi-class IoT attack detection (GPT-2, LLaMA-3.2-1B, Meta-LLaMA-3-8B, and Mistral-v0.3-7B).
\item \textbf{Structured-to-text conversion:} Reformulating numerical features into concise natural language prompts.
\item \textbf{Known and unknown attack generalization:} Leveraging RAG to enable zero-shot detection beyond traditional supervised models.
\item \textbf{Comparative evaluation:} Benchmarking LLMs against classical ML classifiers using standard metrics.
\end{itemize}
The paper is organized as follows. Section~\ref{sec:related} reviews intrusion detection from classical ML to modern LLMs. Section~\ref{sec:method} details our methodology, including data preprocessing, QLoRA fine-tuning, and RAG. Section~\ref{sec:results} evaluates performance on known and unseen attacks. Finally, Section~\ref{sec:conclusion} concludes and outlines limitations and future work.

\section{Related Work} \label{sec:related}

This section surveys the evolution of network intrusion detection, from classical machine learning to contemporary LLM-based methods, culminating in the challenge of detecting previously unseen attacks.

Classical machine learning techniques, such as Random Forest (RF) and Support Vector Machines (SVM), have been extensively employed for intrusion detection, demonstrating strong performance on balanced datasets~\cite{chang2017network,hiremath2025ml}. Nevertheless, their efficacy diminishes in multi-class scenarios with imbalanced distributions~\cite{disha2022giwrf}. Deep learning approaches, particularly Convolutional Neural Networks (CNNs), have been explored to capture spatial and temporal patterns in network traffic, achieving improved detection accuracy~\cite{ferrag2020deep}. However, these supervised methods generally cannot identify novel attack types without complete retraining.

The advent of Large Language Models (LLMs) has introduced advanced contextual reasoning capabilities in cybersecurity. Transformer-based architectures can effectively process structured network data transformed into textual prompts~\cite{kheddar2025transformers,kaur2025harnessing}. Parameter-efficient fine-tuning methods, notably QLoRA~\cite{Dettmers2023QLoRA}, facilitate the adaptation of large models under computational constraints, enabling effective fine-tuning of decoder-only LLMs for known attack classification~\cite{houssel2024explainable}.

Detecting zero-day attacks remains a significant challenge. Meta-learning approaches allow models to adapt to new classes with minimal examples~\cite{alrayes2025_maml}, while Generative Adversarial Networks (GANs) have been employed to synthesize data for rare attack types~\cite{shirazi2023_ganzero}. Few-shot and zero-shot learning strategies provide evaluation mechanisms for novel attacks~\cite{althiyabi2024fewshot}, but typically require specialized architectures or extensive retraining. Recent deep-learning IDS frameworks have addressed zero-day and concept-drift scenarios via adaptive, multi-agent architectures~\cite{zaki2024hybrid}. In contrast, our methodology leverages LLMs with retrieval-augmented reasoning to capture semantic relationships across attack behaviors, enabling zero-shot generalization without modifying the model architecture or performing incremental updates.

Retrieval-Augmented Generation (RAG) has emerged as an effective mechanism to incorporate external knowledge, offering contextual evidence for tasks such as log anomaly detection without requiring parameter updates~\cite{pan2023raglog,gao2023retrieval}. Complementarily, hybrid architectures combining statistical detection with LLM-based reasoning have demonstrated improved detection performance in IoT environments~\cite{alhammouri2025_hybrid_iot,loumachi2025gendfir}.

Building on these advances, our approach integrates QLoRA-based fine-tuning for known attacks and employs a RAG framework for previously unseen attacks, providing a unified system that supports both supervised classification and zero-shot generalization in complex IoT networks.

\section{Methodology}
\label{sec:method}

\begin{figure*}[h]
\centering
    \includegraphics[width=0.7\linewidth]{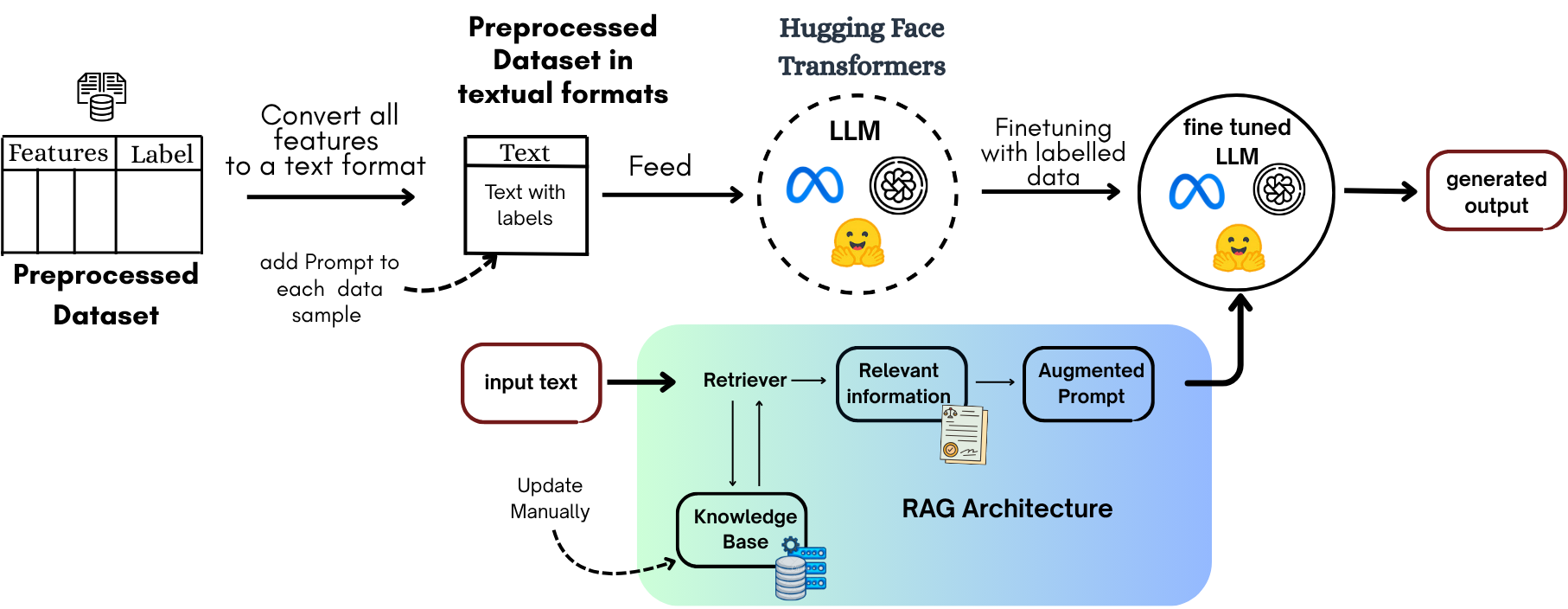}
    \caption{Workflow summary chart}
    \label{fig:methodology1}
\end{figure*}

This study employs an end-to-end methodology comprising dataset preparation, preprocessing, feature selection, structured-to-text conversion, and fine-tuning of lightweight decoder-only LLMs. The workflow integrates prompt design, tokenization, and parameter-efficient adaptation using QLoRA. Traditional machine learning models were implemented as baselines for performance comparison. Furthermore, a Retrieval-Augmented Generation (RAG) framework was incorporated to enable zero-shot detection of unseen attack types. The overall workflow of the proposed approach is illustrated in Fig.~\ref{fig:methodology1}.

\subsection{Data Preparation}

The CICIoT2023 dataset~~\cite{hammal2023ciciot} was used for this study, which comprises network traffic from 105 IoT devices and includes 34 classes (33 attacks and benign). To address the significant class imbalance, a balanced subset was created via random downsampling. From the dataset, 500 samples per class were allocated for model development (training and validation), while a separate hold-out set of 100 samples per class was reserved for final testing. The development set was further divided using an 80/20 train-validation split, a standard practice in IoT IDS research to manage class imbalance in CIC-based benchmarks. A summary of the data splits is provided in Table~\ref{tab:splitting_data}.

\begin{table*}[h]
\setlength{\tabcolsep}{3pt}
\centering
\caption{Data preparation split for both experiments}
\label{tab:splitting_data}
\begin{tabular}{l|c|c|c|c}
\textbf{Dataset Split} & \textbf{Per-Class Count} & \textbf{Split Ratio} & \textbf{Classes} & \textbf{Usage} \\ \hline
\multirow{2}{*}{LLM: Training + Validation} & \multirow{2}{*}{500} & 80\% & 24 & Training subset \\ \cline{3-5}
& & 20\% & 24 & Validation subset \\ \hline
LLM: Testing & 100 & – & 24 & Model evaluation \\ \hline
\multirow{2}{*}{RAG Knowledge Base + RAG Test} & \multirow{2}{*}{1,000} & 70\% & 10 & RAG Knowledge Base (unseen) \\ \cline{3-5}
& & 30\% & 10 & RAG Test (unseen) \\
\end{tabular}
\end{table*}

For RAG experiments, 1,000 samples per class from 10 excluded attack types (unseen classes) were partitioned 70/30 for the retrieval knowledge base and testing, ensuring fair representation across training, retrieval, and evaluation.

\subsection{Data Preprocessing and Feature Engineering}
\subsubsection{Feature Selection}
The  dataset contains 44 numerical features extracted from network flows. To reduce redundancy and multicollinearity, correlation-based feature selection was applied. Features with a Pearson coefficient above 0.98 were considered highly correlated; one feature from each correlated pair was removed manually after following the exploratory data analysis. The final selected 23 features are listed in Table~\ref{tab:selected_features}.

\begin{table}[h]
\setlength{\tabcolsep}{3pt}
\centering
\caption{Feature Descriptions}
\label{tab:selected_features}
\begin{tabular}{l|l}

\textbf{Feature Name} & \textbf{Description} \\
\hline
Header\_Length & Packet header length \\
Protocol Type & Transport/network layer protocol \\
Time\_To\_Live & IP packet TTL value \\
psh\_flag\_number & Packets with PSH flag set \\
ack\_flag\_number & Packets with ACK flag set \\
ack\_count & Acknowledged packets in flow \\
syn\_count & Packets with SYN flag set \\
fin\_count & Packets with FIN flag set \\
rst\_count & Packets with RST flag set \\
HTTP & HTTP usage in flow (binary) \\
HTTPS & HTTPS usage in flow (binary) \\
DNS & DNS usage in flow (binary) \\
TCP & TCP transport protocol (binary) \\
UDP & UDP transport protocol (binary) \\
ICMP & ICMP network protocol (binary) \\
Tot sum & Total packet length in flow \\
Min & Min. packet size in flow \\
Max & Max. packet size in flow \\
AVG & Avg. packet size in flow \\
Std & Std. dev. of packet sizes in flow \\
IAT & Inter-arrival time of packets \\
Number & Total packets in flow \\
Rate & Packet transmission rate in flow \\
Label & Attack/benign class label \\

\end{tabular}
\end{table}

\subsubsection{Data Normalization}
Traditional ML models such as RF and SVM can be sensitive to differences in feature scales. To avoid this issue, Classical ML models were trained on normalized features using Scikit-learn’s \texttt{StandardScaler}, standardizing each feature to zero mean and unit variance. 
For the LLM-based models, normalization was not applied since features were converted to natural language text. All feature values were represented as numerical  values rounded to six decimal places (numerical text tokens).

\subsection{Transformation for LLM Processing and RAG integration}

To adapt the structured network traffic data for LLMs, several preprocessing steps were applied to ensure efficient tokenization and meaningful text representation. The primary objective was to transform tabular features into concise, semantically coherent natural language prompts suitable for fine-tuning.

\subsubsection{Feature and Class Label Standardization}

\begin{table}[h]
\centering
\caption{Feature and Class Label Standardization Examples}
\label{tab:standardization_examples}
\small
\begin{tabular}{|p{0.95\linewidth}|}
\hline
\textit{\textbf{Feature Name Examples:}} \\
Header\_Length $\rightarrow$ Header Length \\
rst\_count $\rightarrow$ Packets with RST Flag \\
IAT $\rightarrow$ Time Between Packets \\ \hline
\textit{\textbf{Class Label Examples:}} \\
DDOS\-UDP\_FLOOD $\rightarrow$ ddos udp flood \\
MITM\-ARPSPOOFING $\rightarrow$ mitm arp spoofing \\
DDOS\-SYN\_FLOOD $\rightarrow$ ddos synchronize flood \\
\hline
\end{tabular}
\end{table}

Feature and class names were standardized to maintain consistency and minimize unnecessary tokenization. Special characters such as hyphens and underscores were replaced with spaces, reducing input sequence length without affecting semantic clarity. Class labels were also simplified into meaningful and LLM-friendly names to ensure alignment with the model’s vocabulary. Representative  three examples from features and classes are shown in Table~\ref{tab:standardization_examples}, and the full set of selected seen/unseen classes is listed in Table~\ref{tab:seen_unseen_classes}.

\begin{table}[ht!]
\small
\setlength{\tabcolsep}{4pt}
\centering
\caption{Seen and Unseen Attack Classes Used in the Study}
\label{tab:seen_unseen_classes}

\textbf{Seen Classes (Training/Validation)} \\[3pt]

\begin{tabularx}{\linewidth}{X X}
\hline
benign & ddos ack fragmentation \\
ddos icmp flood & ddos icmp fragmentation \\
ddos pshack flood & ddos rst fin flood \\
ddos synchronize flood & ddos synonymousip flood \\
ddos tcp flood & ddos udp flood \\
ddos udp fragmentation & dns spoofing \\
dos http flood & dos synchronize flood \\
dos tcp flood & dos udp flood \\
mirai greeth flood & mirai greip flood \\
mirai udp plain & mitm arp spoofing \\
recon host discovery & recon os scan \\
recon port scan & vulnerability scan \\
\end{tabularx}

\textbf{Unseen Classes (RAG Evaluation)} \\[3pt]
\begin{tabularx}{\linewidth}{X X}
\hline
backdoor malware & browser hijacking \\
command injection & ddos http flood \\
ddos slow loris & dictionary brute force \\
recon ping sweep & sql injection \\
cross site scripting & uploading attack \\
\end{tabularx}
\end{table}

\subsubsection{Prompt Engineering and Tokenization}
Each record was transformed into a structured prompt containing the task description, possible attack classes and feature–value pairs. An example is shown in Table~\ref{tab:llm_example}. During training, the ground-truth label followed the \texttt{Answer:} token and during the inference, the model predict the class after this token. 

\begin{table}[h]
\centering
\caption{Example Prompt with generated output in bold text}
\label{tab:llm_example}
\small
\begin{tabular}{|p{0.95\linewidth}|}
\hline
\textit{Task: Network Attack Classification} \\
\textit{Input Features: \{Header Length = 20.0; Protocol Type = 6.0; IP Time to Live = 64.0; Flow Packet Transmission Rate = 41913.7; \dots\}} \\
\textit{Possible Classes: [benign, ddos icmp flood, browser hijacking, backdoor, \dots]} \\
\textit{Answer: \textbf{benign}} \\
\hline
\end{tabular}
\end{table}

Tokenization was performed using Hugging Face’s \texttt{AutoTokenizer} with truncation to 512 tokens. Since decoder-only models lack a default padding token, the end-of-sequence (\texttt{<eos>}) token was used for padding. A custom data collator extended from \texttt{DataCollatorForLanguageModeling} ensured that only label tokens (after \texttt{Answer:}) contributed to loss computation, focusing optimization solely on predicting the correct class.

\subsubsection{Parameter-Efficient Fine-Tuning with QLoRA}
\label{Qlora}
For the LLM fine tuning, the QLORA technique was used.  QLoRA combines 4-bit quantization with Low-Rank Adaptation to fine-tune large decoders efficiently under resource constraints. Model weights are loaded in 4-bit precision with mixed FP16 computation, double quantization, and NF4 for numerical stability. LoRA adapters are applied to attention projections with rank $r=16$, scaling factor $\alpha=32$, dropout 0.1, and frozen biases. Most parameters remain frozen, so only a small subset is trainable, reducing memory and computational requirements while maintaining performance across all LLMs.

\subsubsection{Generalization to Unseen Attack Types with RAG}
To evaluate model generalization, a Retrieval-Augmented Generation (RAG) framework was integrated into the fine-tuned LLM workflow. Attack types excluded from training were used as unseen classes. Their feature vectors were embedded to construct a retrieval knowledge base. During inference, each test instance was compared against this base using cosine similarity, and the top 20 most similar samples were identified. The top 3 retrieved examples were concatenated with the query instance (within token constraints) to form the final prompt.

This prompt provided contextual reference for the LLM to infer unseen attack types. The model’s predictions were generated based on both the query flow and the retrieved exemplars. Evaluation metrics included Accuracy, Precision, Recall, and F1-score for classification, and top-$k$ (k=3) recall for retrieval effectiveness. This methodological setup enabled structured assessment of zero-shot generalization capability using retrieval support.

\subsubsection{Hardware Setup}
LLM experiments were conducted on two separate GPUs: NVIDIA RTX 4080 (16\,GiB CUDA memory) for GPT-2 and LLaMA-3.2-1B, and NVIDIA RTX 4090 (32\,GiB CUDA memory) for Mistral-7B and LLaMA-3-8B fine-tuning.

\section{Results}
\label{sec:results}

\begin{figure*}[t!]
    \centering
    \includegraphics[width=0.95\linewidth]{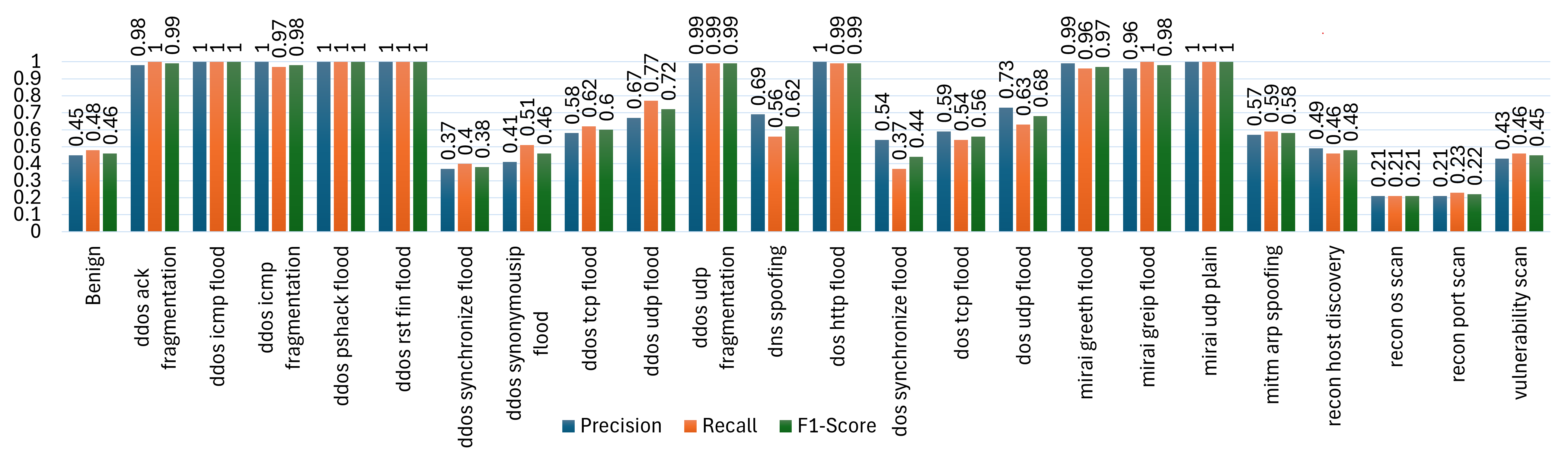}
    \caption{Per-Class Precision, Recall, and for LLaMA 3.2-1B }
    \label{fig:bar}
\end{figure*}
This section presents the outcomes of the two experiments:
 (1) direct classification using fine-tuned models and classical
 baselines, and (2) RAG-based evaluation for unseen attacks.

\subsection{Experiment 1: Direct Classification}

\subsubsection{Model Training and Hyperparameters}
In the first phase, the models are trained and evaluated on attack classes included in the training set. Classical ML classifiers (RF, SVM, and Logistic Regression (LR)) are trained directly on normalized feature vectors. In parallel, decoder-based LLMs including GPT-2, LLaMA-3.2-1B, Meta-LLaMA-3-8B, and Mistral-v0.3-7B are fine-tuned on textual prompts constructed from the same features. All LLM models were fine-tuned using the QLoRA setup described earlier.
The main fine-tuning settings are summarized in Table~\ref{tab:llm_config}.

\begin{table}[b!]
\setlength{\tabcolsep}{3pt}
\centering
\caption{LLM Fine-tuning Configuration}
\label{tab:llm_config}
\begin{tabular}{l|l}
\textbf{Parameter} & \textbf{Value} \\ \hline
LoRA Settings & Rank $r=16$, Scaling $\alpha=32$, Dropout = 0.1 \\ 
Target Layers & Attention/ MLP \\ 
Epochs & 3--5 \\ 
Batch Size & 4 (7B and 8B model),16 (other models)\\ 
Learning Rate & varies (0.0005 or 0.00005) \\ 
Optimizer & AdamW \\ 
Weight Decay & 0.2 \\ 
Mixed Precision & FP16 \\ 
\end{tabular}
\end{table}

\subsubsection{Parameter Efficiency}
Table~\ref{tab:loraparams} reports the number and proportion of trainable parameters. Only a small fraction of each model’s parameters are updated, confirming QLoRA’s parameter efficiency and suitability for resource-limited deployment.

\begin{table}
    \setlength{\tabcolsep}{3pt}
    \centering
    \caption{Trainable Parameters Compared to Total Model Size}
    \label{tab:loraparams}
    \begin{tabular}{l|c|c|l}
        \textbf{Model} & \textbf{Trainable} & \textbf{Total} & \textbf{Trainable \%} \\ \hline
        GPT-2 & 1.57M & 356.40M & 0.44\% \\
        LLaMA-3.2 1B & 2.36M & 1,238.17M & 0.19\% \\
        Mistral-v0.3-7B & 9.44M & 7,257.46M & 0.13\% \\
        LLaMA-3.1 8B & 9.44M & 8,039.70M & 0.12\% \\
    \end{tabular}
\end{table}

\subsubsection{Performance Evaluation}
Model performance was evaluated using Accuracy, Precision, Recall, and F1-score on the test dataset. Results are summarized in Table~\ref{tab:test_results_runtime}.

\begin{table}[h]
    \setlength{\tabcolsep}{3pt}
    \centering
    \caption{Test Set Evaluation Metrics for All Models}
    \label{tab:test_results_runtime}
    \begin{tabular}{l|c|c|c|c|c}
    \textbf{Model} & \textbf{Accuracy} & \textbf{F1} & \textbf{Precision} & \textbf{Recall} & \textbf{Runtime/s} \\
    \hline
    LR & 0.6792 & 0.6778 & 0.6792 & 0.6676 & 0.0004 \\
    RF & \textbf{0.7171} & \textbf{0.7159} & 0.7171 & 0.7143 & 0.0245 \\
    SVM & 0.6470 & 0.6761 & 0.6650 & 0.6650 & 0.6718 \\\hline
    LLaMA-3.2-1B & 0.7117 & 0.7124 & 0.7173 & 0.7117 & 235.16 \\
    Mistral-7B & 0.7104 & 0.6992 & 0.7210 & 0.7104 & 1050.58 \\
    LLaMA-3.8-8B & 0.6796 & 0.6792 & 0.6824 & 0.6796 & 679.51 \\
    GPT-2 & 0.6567 & 0.6271 & 0.6617 & 0.6567 & 191.05 \\
    \end{tabular}
\end{table}

Among baseline models, \textit{Random Forest} achieved the highest performance (Accuracy = 0.7171, F1 = 0.7159). For fine-tuned LLMs, LLaMA 3.2-1B achieved the best F1-score (\textbf{0.7124}) with balanced metrics, followed by Mistral-7B (F1 = 0.6992). Although LLMs require higher inference time, their runtime remains practical for offline or batch intrusion analysis. LLaMA-3.2-1B offers a favorable balance between accuracy and compute cost (235 s), showing that lightweight LLMs can deliver strong detection under modest hardware while enabling zero-shot generalization not achievable with traditional models.

\subsubsection{Per-Class Performance}
LLaMA 3.2-1B, the best-performing fine-tuned model, was used for detailed class-wise evaluation. Results are presented in Table~\ref{tab:classification_report}.

\begin{table}[htbp]
\setlength{\tabcolsep}{3pt}
\centering
\caption{Per-Class Classification Report for Test Set (LLaMA 3.2-1B)}
\label{tab:classification_report}
\renewcommand{\arraystretch}{1.1}
\begin{tabular}{l|c|c|c|c}
\textbf{Class} & \textbf{Precision} & \textbf{Recall} & \textbf{F1} & \textbf{Support} \\
\hline
Benign & 0.45 & 0.48 & 0.46 & 100 \\
ddos ack fragmentation & 0.98 & 1.00 & 0.99 & 100 \\
ddos icmp flood & 1.00 & 1.00 & 1.00 & 100 \\
ddos icmp fragmentation & 1.00 & 0.97 & 0.98 & 100 \\
ddos pshack flood & 1.00 & 1.00 & 1.00 & 100 \\
ddos rst fin flood & 1.00 & 1.00 & 1.00 & 100 \\
ddos synchronize flood & 0.37 & 0.40 & 0.38 & 100 \\
ddos synonymousip flood & 0.41 & 0.51 & 0.46 & 100 \\
ddos tcp flood & 0.58 & 0.62 & 0.60 & 100 \\
ddos udp flood & 0.67 & 0.77 & 0.72 & 100 \\
ddos udp fragmentation & 0.99 & 0.99 & 0.99 & 100 \\
dns spoofing & 0.69 & 0.56 & 0.62 & 100 \\
dos http flood & 1.00 & 0.99 & 0.99 & 100 \\
dos synchronize flood & 0.54 & 0.37 & 0.44 & 100 \\
dos tcp flood & 0.59 & 0.54 & 0.56 & 100 \\
dos udp flood & 0.73 & 0.63 & 0.68 & 100 \\
mirai greeth flood & 0.99 & 0.96 & 0.97 & 100 \\
mirai greip flood & 0.96 & 1.00 & 0.98 & 100 \\
mirai udp plain & 1.00 & 1.00 & 1.00 & 100 \\
mitm arp spoofing & 0.57 & 0.59 & 0.58 & 100 \\
recon host discovery & 0.49 & 0.46 & 0.48 & 100 \\
recon os scan & 0.21 & 0.21 & 0.21 & 100 \\
recon port scan & 0.21 & 0.23 & 0.22 & 100 \\
vulnerability scan & 0.43 & 0.46 & 0.45 & 100 \\
\hline
\textbf{Macro Avg} & 0.70 & 0.70 & 0.70 & 2400 \\
\textbf{Weighted Avg} & 0.70 & 0.70 & 0.70 & 2400 \\
\hline
\textbf{Accuracy} & \multicolumn{3}{c|}{0.70} & 2400 \\
\end{tabular}
\end{table}

High F1-scores ($\geq0.98$) for \textit{DDoS} and \textit{Mirai}-based attacks such as \textit{ddos icmp flood}, \textit{mirai udp plain}, and \textit{ddos pshack flood} indicate strong detection of highly distinctive attack signatures. Conversely, lower F1-scores for \textit{recon os scan}, \textit{vulnerability scan}, and \textit{ddos synchronize flood} reveal difficulty in differentiating subtle or overlapping network behaviors. 

The per-class metric plot (Fig.~\ref{fig:bar}) visually reinforces the quantitative findings, highlighting strong predictive accuracy for most attack types while revealing misclassifications in classes that share similar traffic characteristics.

\subsection{Experiment 2: RAG-Based Evaluation for Unseen Attacks}

\subsubsection{Numerical Embedding Retrieval}
Before passing input prompts to the LLM, we evaluated the ability of the numerical embeddings to retrieve relevant examples from the RAG knowledge base. For each test instance, we computed cosine similarity against all examples in the knowledge base and considered the top 20 candidates. The metric used is Top-3 Recall, which indicates whether at least one of the top 3 retrieved candidates matches the actual class of the test sample. Per-Class evaluation is shown in Table \ref{tab:rag_per_class}.

Across all 3,000 test samples, the overall Top-3 Recall was 63.27\% (1898/3000), indicating that for nearly two-thirds of the test instances, at least one of the top 3 candidates retrieved from the knowledge base matched the true class. This retrieval step ensures that the subsequent LLM input is enriched with relevant examples, improving its ability to classify unseen or difficult attack types.

\subsubsection{RAG Prompt Construction and Generation Settings}
From the top 20 retrieved candidates, the three most similar were concatenated with the current input. Prompts were truncated to 1015 tokens and limited to six generated tokens. The final prediction was extracted after the last \texttt{Answer:} token. The Table~\ref{tab:rag_examples} illustrates a RAG-augmented prompt example used and its model output (shown in bold).








\begin{table}[h]
\centering
\caption{RAG-augmented prompt example with generated output in bold text}
\label{tab:rag_examples}
\small
\begin{tabular}{|p{0.95\linewidth}|}
\hline
Retrieved Examples: \\ 
Input: \{Header Length=27.6; Protocol Type=6.0; IP TTL=57.3; \dots; Total Packets=10\} \\Answer: recon ping sweep \\
Input: \{Header Length=25.6; Protocol Type=6.0; IP TTL=78.8; \dots; Total Packets=10\} \\Answer: sql injection \\
Input: \{Header Length=21.6; Protocol Type=6.0; IP TTL=56.8; \dots; Total Packets=10\} \\Answer: recon ping sweep \\

Task: Network Attack Classification; \\Possible Classes: [recon ping sweep, sql injection] \\
Example Input: \{Header Length=25.6; Protocol Type=6.0; IP TTL=52.8; \dots; Total Packets=10\} \\Answer: \textbf{recon ping sweep} \\
\hline
\end{tabular}
\end{table}

This format helps the model reason about unseen attacks by grounding predictions in semantically similar examples.

\subsubsection{RAG-Based Classification Performance}

Per-class evaluation metrics for the RAG-based experiment are presented in Table~\ref{tab:rag_class_report}. The fine-tuned model achieved an overall accuracy of 42.63\% on previously unseen attack classes, demonstrating its capacity for zero-shot generalization without any additional supervised retraining. 

\begin{table}[h!]
\setlength{\tabcolsep}{3pt}
    \centering
    \caption{Top-3 Recall (\%) per Class Using Numerical Embeddings}
    \label{tab:rag_per_class}
    \begin{tabular}{l|c}
    \textbf{Attack Class} & \textbf{Top-3 Recall (\%)} \\ \hline
    Recon Ping Sweep & 59.59 (174/292) \\
    DDoS Slow Loris & 99.06 (317/320) \\
    Browser Hijacking & 59.09 (182/308) \\
    Backdoor Malware & 41.19 (131/318) \\
    Dictionary Brute Force & 52.76 (153/290) \\
    Command Injection & 60.42 (174/288) \\
    SQL Injection & 59.25 (173/292) \\
    DDoS HTTP Flood & 97.36 (295/303) \\
    Cross Site Scripting & 51.70 (137/265) \\
    Uploading Attack & 50.00 (162/324) \\
    \end{tabular}
\end{table}

\begin{table}[h!]
\setlength{\tabcolsep}{3pt}
\centering
\caption{Per-Class Results for RAG-Based Evaluation}
\label{tab:rag_class_report}
\begin{tabular}{l|c|c|c|c}
\textbf{Class} & \textbf{Precision} & \textbf{Recall} & \textbf{F1} & \textbf{Support} \\ \hline
Recon Ping Sweep & 0.35 & 0.35 & 0.35 & 292 \\
DDoS Slow Loris & 0.96 & 0.94 & 0.95 & 320 \\
Browser Hijacking & 0.39 & 0.35 & 0.37 & 308 \\
Backdoor Malware & 0.24 & 0.21 & 0.22 & 318 \\
Dictionary Brute Force & 0.29 & 0.31 & 0.30 & 290 \\
Command Injection & 0.33 & 0.38 & 0.35 & 288 \\
SQL Injection & 0.30 & 0.28 & 0.29 & 292 \\
DDoS HTTP Flood & 0.94 & 0.96 & 0.95 & 303 \\
Cross Site Scripting & 0.16 & 0.20 & 0.18 & 265 \\
Uploading Attack & 0.26 & 0.24 & 0.25 & 324 \\
\hline
\textbf{Macro Avg} & 0.38 & 0.38 & 0.38 & 2990 \\
\textbf{Weighted Avg} & 0.43 & 0.43 & 0.43 & 2990 \\
\hline
\textbf{Accuracy} & \multicolumn{3}{c|}{0.426} & 2990\\ 
\end{tabular}
\end{table}

The results in Table~\ref{tab:rag_class_report} indicate that RAG-enhanced inference substantially improves the LLM’s ability to recognize semantically distinct attack types. High precision and recall scores for classes such as \textit{DDoS Slow Loris} and \textit{DDoS HTTP Flood} suggest that the retrieval mechanism effectively supplements contextual understanding for well-represented behavioral patterns in the external knowledge base. In contrast, lower performance on complex or less distinctive attacks like \textit{Backdoor Malware}, \textit{Cross Site Scripting}, and \textit{Uploading Attack} reflects the inherent challenge of modeling subtle variations in traffic features when semantic overlap exists between categories.  

These findings demonstrate that integrating retrieval-based augmentation helps bridge the gap between seen and unseen classes, enabling LLMs to extend beyond fixed supervised boundaries typical of classical ML models. Although overall accuracy remains moderate, the RAG framework shows promise for incremental adaptation, where the knowledge base can be continuously enriched with new attack exemplars to improve coverage and robustness against emerging IoT threats. This suggests a viable direction for future work toward hybrid retrieval–generation intrusion detection frameworks.


\section{Conclusions and Future Work}
\label{sec:conclusion}

This study shows that lightweight decoder-only LLMs, adapted using structured-to-text conversion, QLoRA fine-tuning, and RAG, form an efficient unified framework for IoT intrusion detection. On CICIoT2023, the fine-tuned LLaMA-1B model achieved an F1-score of 0.7124, matching the Random Forest baseline (0.7159) on known attacks, while the RAG-enhanced model reached 42.63\% accuracy on unseen attack types without additional training, demonstrating practical zero-shot capability beyond traditional ML. By combining parameter efficiency with retrieval-based context grounding, the proposed approach offers a scalable and resource-aware solution suitable for evolving IoT environments. A key limitation is that results were obtained on a single dataset, and broader validation across heterogeneous IoT benchmarks is required. Future work will explore improved retrieval strategies, larger and more diverse knowledge bases, and lightweight ensemble methods to enhance detection of subtle or overlapping attack patterns.

\bibliographystyle{IEEEtran}
\bibliography{references}


\end{document}